\documentstyle[aps,prb,epsf]{revtex}
\preprint{IASSNS-HEP-00/02}
\date{\today}
\title{\strut{\vadjust{\vskip-0.5in\strut\hfill{\normalsize
        IASSNS-HEP-00/02\vskip0.25in}}} %
Holons on a meandering stripe: quantum numbers}
\author{
Oleg Tchernyshyov\cite{email-oleg} and Leonid P. Pryadko}
\address{School of Natural Sciences, Institute for Advanced Study, 
Princeton, New Jersey 08540}

\begin{document}
\twocolumn[\hsize\textwidth\columnwidth\hsize\csname %
  @twocolumnfalse\endcsname

\maketitle
\begin{abstract}
We attempt to access the regime of strong coupling between charge
carriers and transverse dynamics of an isolated conducting ``stripe'',
such as those found in cuprate superconductors.  A stripe is modeled
as a partially doped domain wall in an antiferromagnet (AF),
introduced in the context of two different models: the $t$--$J$ model
with strong Ising anisotropy, and the Hubbard model in the
Hartree-Fock approximation.  The domain walls with a given linear
charge density are supported artificially by boundary conditions.  In
both models we find a regime of parameters where doped holes lose
their spin and become holons (charge $Q=1$, spin $S_3=0$), which can
move along the stripe without frustrating AF environment.  One aspect
in which the holons on the AF domain wall differ from those in an
ordinary one-dimensional electron gas is their transverse degree of
freedom: a mobile holon always resides on a transverse kink (or
antikink) of the domain wall.  This gives rise to two holon flavors
and to a strong coupling between doped charges and transverse
fluctuations of a stripe.
\end{abstract}
]
\narrowtext
\newcommand{\lsco}{La$_{2-x}$Sr$_x$CuO$_4$\/}
\newcommand{\beq}{\begin{equation}}
\newcommand{\eeq}{\end{equation}}
\newcommand{\beqa}{\begin{eqnarray}}
\newcommand{\eeqa}{\end{eqnarray}}

\section{Introduction}
\label{Sec_Introduction}

The nature of charge carriers in high-$T_c$ cuprate superconductors
remains a subject of debate.  The stoichiometric (``parent'')
compounds are antiferromagnetic (AF) insulators well described by the
Heisenberg model.  Experiments show that, at least in some of the
cuprates, doping with holes creates an intrinsically inhomogeneous
state with periodically modulated charge and staggered spin densities.
Neutron and X-ray scattering experiments indicate that staggered
magnetization has a period of modulation that is twice as long as that
of charge density.\cite{neutron,x-ray} This is consistent with the
notion of charged stripes separating AF domains with alternating Neel
magnetization.\cite{Tranquada}

Stripes as domain walls in the ground state of a collinearly ordered
antiferromagnet were predicted---prior to reliable experimental
detection---on the basis of Hartree-Fock studies of the Hubbard
model\cite{early-HF} near half-filling.  Mean-field calculations yield
a linear density of $\nu=1$ doped hole per lattice site, which almost
certainly means insulating stripes, in apparent contradiction with
experiments.  Besides, stripes observed in the cuprates tend to have a
linear hole density $\nu \approx 1/2$, at least when they are
sufficiently well separated.\cite{Yamada} To date, no reliable
microscopic calculation yields the experimentally observed filling
fraction~$\nu$, let alone explains the transport and high-temperature
superconductivity in the cuprates.  Numerical simulations have so far 
been inconclusive.\cite{White-Scalapino,Hellberg-Manousakis}

In the absence of a reliable microscopic theory, attempts have been
made to find a phenomenological description of the stripes.  In one of
the more popular routes, a stripe is modeled\cite{EK} as a
one-dimensional electron gas (1DEG) interacting with the surrounding
environment\cite{Zachar-Kivelson} and with the transverse motion of
the stripe.\cite{Kivelson-Fradkin,Senthil} As we argue below, this
approach rests on the assumption that the physics of an isolated
stripe is basically the same as in the limit~$\nu\to 1$.  In this
limit, a stripe is almost completely filled with holes and can be
described as an electron gas at low density $1-\nu$.  It is far from
obvious, though certainly not implausible, that stripes with
$\nu\approx 1$ and $\nu\approx 1/2$ should exhibit qualitatively
similar behavior.

In this work we develop 
a qualitatively different (but not less plausible) phenomenology of a
partially doped stripe.  It is based on two different model
calculations performed in the limit of low {\em hole\/} density on a
stripe, $\nu\ll1$.  Nominally, this is as far from the observed
density $\nu\approx1/2$ as the electron-gas limit $\nu\to1$.  The
quantum numbers of charge carriers (holons) in our model calculations
are completely different from those of electrons.  It likely means
that the two limits are not adiabatically connected.  It is clear then
that the phase $\nu\approx1/2$ can resemble only one of the
low-density limits: either $\nu\to 1$, %
or $\nu\to 0$---or possibly none of the above!

Building on our model calculations we conjecture that charge
carriers of the ``$\nu\to0$'' phase are {\em holons} (charge $Q=1$,
spin $S_3=0$).  In both models the loss of spin is compensated by the
emergence of another spin-like degree of freedom, termed the
transversal flavor by Zaanen {\em et al.}\cite{Zaanen9804300} This
happens because a holon always resides on a transverse kink or
antikink of a domain wall.  Thus holons are strongly coupled
to transverse fluctuations of a stripe.  Yet, the motion of such
objects along the stripe is free, it does not produce any additional
spin frustration.  Such a holon gas is clearly very different from the
electron gas of the ``$\nu\to1$'' phase.

Of course, our approach should not be interpreted as a suggestion
that a stripe with small linear hole density $\nu\ll1$ can be stable
in any model relevant to high-$T_c$ materials.  On the contrary, a
domain wall in an undoped antiferromagnet is a highly excited texture.
A finite linear density of holes is needed to stabilize a domain wall.
In our analysis we always assume that the domain wall is supported
externally ({\em e.g.}, by the boundary conditions), while its
untwisting is suppressed by a sufficiently strong anisotropy (strictly
linear polarization in our Hartree-Fock analysis).  In a model where
partially filled stripes appear {\em in the ground state}, neither
assumption would be necessary.

Indeed, we have reduced the symmetry of the problem in order to
stabilize topologically the Ising-type domain
walls.\cite{Kivelson-symmetry} In any model with the O(3) symmetry of
the Neel order parameter, there can be no topological arguments for
their stability.  For example, domain walls can be continuously
untwisted in a broad class of Ginzburg-Landau models with a continuous
O$(N)$ symmetry; such domain walls are {\em locally unstable\/}.
However, local instabilities are not an issue for globally stable
configurations.  In practical terms, untwisting does not occur if
domain walls appear in the ground state of a system.  In the context
of high-$T_c$ stripe phases, such nontopologically-stable domain walls
were discussed in Ref.~\CITE{Pryadko-MF}.  Their global stability
requires frustration of the AF order on some microscopic or
intermediate length scale, e.g., as a result of doping.

Therefore, our model calculations should be viewed as an attempt
to identify plausible ground states of an isolated stripe.  In
contrast to phenomenological approaches, we do not postulate effective
one-dimensional models of a stripe.  Instead, we derive them by
starting with a two-dimensional model describing an antiferromagnet
with a domain wall.  Our 2D models may be unrealistic for the
cuprates, but the resulting 1D effective theories have the set of
elementary excitations consistent with the paradigm of a stripe as a
doped fluctuating domain wall in an antiferromagnet.  In essence, we
rely on universality: if the number of qualitatively different ground
states of a stripe (classified by quantum numbers and spectrum of
low-lying excitations\cite{EKZ-99}) is limited, all of them
may be derived from simple 2D models.

From this perspective, our work adds the 1D holon gas to the list of
potential stripe models.  Note, however, that, in the presence of
interactions, the description of a stripe in terms of holons, and a
more conventional one in terms of electrons, are not necessarily
incompatible.  Both models can be viewed as Luttinger liquids with
different collective modes: charge and spin in the case of electrons,
charge and transverse fluctuations for
holons.\cite{Pryadko-Tchernyshyov} Thus, 1D electrons with a spin gap
and 1D holons with a transverse gap may well represent one and the
same phase.  We intend to discuss the role of interactions among the
holons in a future publication.

The paper is organized as follows.  In Sec.~\ref{Sec-t-J} we analyze
partially doped domain walls in a $t$--$J$ model with large Ising
anisotropy.  The ground state and the spectrum of elementary
excitations (spinons and holons) are found explicitly in the limit of
small doping $\nu$.  In Sec.~\ref{Hubbard-numerics} we present
numerical evidence for holons in the Hubbard model, within the
Hartree-Fock approach.  The Hartree-Fock equations for the Hubbard
model are further analyzed in Sec.~\ref{Hubbard-analytics}, where we
introduce an appropriate long-wavelength approximation and study the
spectrum of midgap states induced by a domain wall.  We give heuristic
arguments for the existence of fermion zero modes around a transverse
kink on a domain wall.  Technical details are collected in the
Appendixes.

\section{Holons on a domain wall:
  \hbox{$t$--$J$} model with Ising anisotropy.}
\label{Sec-t-J}

A simple yet very instructive example of holon gas on a domain wall
is offered by the $t$--$J$ model with Ising anisotropy, previously
considered by Kivelson {\em et al.},\cite{Kivelson-tJz} 
\begin{eqnarray}
  \label{t-J-Ham}
  H_{t-J_z} &=&\! \sum_{\langle{\bf rr'}\rangle}\biggl\{\Bigl[ 
    -t\, a^\dagger_{\sigma}({\bf r'}) \, a_{\sigma}({\bf r})
    +%
    {J_\perp\over2}\, s_+({\bf r'})\, s_-({\bf r}) + \,  {\rm h.c.}
    \Bigr] \nonumber \\ 
  & & \qquad\;+\,  J_z\, s_z({\bf r'})\, s_z({\bf r}) 
  + V\, n({\bf r'}) \, n ({\bf r})\biggr\},
\end{eqnarray} 
with the usual constraint of no double occupancies; the sum is taken
over pairs of nearest-neighbor sites.  The $t$--$J$ model proper is
restored if we set $J_\perp=J_z=-4\,V$.  

The analysis of the model~(\ref{t-J-Ham}) is greatly simplified in the
strongly-anisotropic limit $J_z\to\infty$ (while the energy of an AF
bond $E_b = V-J_z/4$ may remain finite).  In this case, an isolated
hole in the AF bulk is localized and has the energy $\epsilon=-4E_b$,
the cost of 4 missing bonds.  Two holes on adjacent sites share one
missing bond, which is to say that their interaction energy is $E_b$.
When severing a bond is costly ($E_b < 0$ and has a large absolute
value), hole-rich islands are formed in the otherwise unaltered
antiferromagnet, with the energy $-2E_b$ per doped hole (phase
separation in the bulk).  In the opposite limit, $E_b \gg |t|$, doped
holes strongly repel one another and stay apart.  The energy per hole
in this case is\cite{Kivelson-tJz}
\begin{equation}
  \label{eq:bulk-localized-hole}
  \varepsilon_{{\rm hole}}=-4\,E_b = J_z-4V. 
\end{equation}
Phase separation or not, doped holes are immobilized in the bulk of an
antiferromagnet.  As long as $J_z$ greatly exceeds both $t$ and $J_\perp$,
the cost of frustrated (ferromagnetic) bonds produced by a moving hole
outweighs the reduction in kinetic energy.  

This situation changes dramatically in the presence of a domain wall:
doped holes become mobile.

\subsection{Spinons on a domain wall}

\begin{figure}
  \centering
  \epsfxsize\columnwidth
  \epsffile{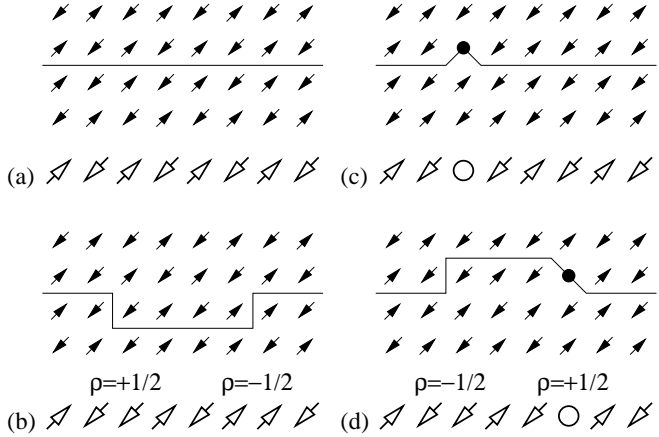}
  \vskip 3mm
  \caption{Elementary excitations of a bond-centered domain wall in the
    anisotropic $t$--$J$ model.  $\rho$ is the transverse flavor.
    (a--b) Two spin exchanges applied to the vacuum state create two
    separated spinons.  (c--d) As a result of hopping, a doped hole
    leaves behind a spinon and becomes a holon.  An antisymmetric spin
    configuration around the holon means that it has $S_z=0$.}
\label{Fig-tJ-spinon}
\end{figure}

Elementary excitations of an undoped domain wall are kinks formed in
pairs by flipping two nearest-neighbor spins [Fig.~\ref{Fig-tJ-spinon}
(a), (b)].  The kinks are mobile, carry zero charge and spin
$s_z=\pm1/2$.  We thus term them {\em spinons\/}.  To make their
relation to 1D spinons\cite{Schulz-spinons} more explicit, we have
integrated spin $s_z$ {\em across} the domain wall with a smooth
envelope to obtain an effective 1D spin chain representing the domain
wall [Fig.~\ref{Fig-tJ-spinon}~(a),~(b), open symbols].  By using Bloch
states,\cite{0-pi} one finds the spinon energy spectrum,
\begin{equation}
  \label{eq:spinon-energy}
  E_{\rm spinon}(k_x) = J_z/2 + J_\perp\,\cos{2 k_x} + {\cal
    O}(J_\perp^2/J_z) > 0. 
\end{equation}
Clearly, for $J_z\gg J_\perp$, such excitations are strongly gapped;
an undoped domain wall is very stiff. 

\subsection{Holons on a domain wall}

A single hole doped into a straight domain wall
[Fig.~\ref{Fig-tJ-spinon}~(c),~(d)] cannot hop along the stripe in the
limit $J_z\to \infty$: it must leave a spinon at the original location
of the hole, which costs ${\cal O}(J_z)$ in magnetic
energy.\cite{endnote:trugman}  Therefore, a single hole can only
oscillate across the stripe; the corresponding ground-state energy is
\begin{equation}
  \label{eq:single-hole-energy}
  E_{\rm hole}=-4\,E_b-{J_z\over2}-|t|+{\cal O}(t^2/J_z).
\end{equation}

It is important to realize that, once a hole starts moving along the
domain wall, it does {\em not\/} create a string of ferromagnetic
bonds, which localize a hole in the bulk.  A moving charge is now
associated with a kink in the transverse position of the stripe
[Fig.~\ref{Fig-tJ-spinon}~(d)].  This composite object has spin
$S_z=0$ and charge $Q=+1$.  Following the spin-chain convention it can
be termed a {\em holon\/}.  Moreover, holons can be created in pairs
without spinons: two spinons created by two holes can annihilate each
other.  We can say that a localized doped hole decays virtually into a
spinon and a holon.  Another hole nearby can absorb the costly spinon
and become a holon.
The intermediate spinon is not needed if the two holes are on adjacent
sites on the same side of the wall.

A holon with momentum $k_x$ along the stripe has the energy
\begin{equation}
  \label{eq:holon-energy}
  E_{\rm holon}(k_x)=-4\,E_b-{J_z\over2}-2|t|\cos{k_x}+{\cal
  O}(t^2/J_z). 
\end{equation}
For small momentum $k_x$, this is smaller then the energy of a single
hole~(\ref{eq:single-hole-energy}).  Therefore, a dilute gas of holons
has a lower energy than a collection of holes similar to that in
Fig.~\ref{Fig-tJ-spinon}~(c).  As shown in
Appendix~\ref{app:lower-bound}, a dilute holon gas is stable against
phase separation for $E_b>0$ when the interaction between the holons
is strictly repulsive.  When $E_b<0$, holes on adjacent sites
attract.  This attraction wins over an increase in kinetic energy for
$E_b\lesssim -t$, causing phase separation: holes can lower their
energy by forming a densely populated ($\nu=1$) stripe leaving the
rest of the domain wall undoped (see Appendix~\ref{app:lower-bound}).

\subsection{Preferred linear charge density}

So far we have considered an antiferromagnet with a single domain wall
maintained by the appropriate boundary conditions.  Within this model,
one can study arbitrary linear concentrations of holes $\nu$ on the
wall.  Particularly simple situations are the limit of dilute holes
$\nu \to 0$ (gas of holons with a transverse flavor) and the opposite
limit $\nu \to 1$ (1D electron gas).

Because $J_z$ is large, partially doped ($\nu<1$) stripes do not occur
naturally in this model.  The preferred linear density of charge $\nu$
can be found by using the usual Maxwell
construction.\cite{Emery-phase-separation} To do so, one minimizes
the energy per doped hole---including the cost of creating domain
walls.  Since partially doped domain walls contain costly
ferromagnetic bonds, they will not occur if $J_z \to \infty$.
A lower bound for the energy per doped hole is
\begin{displaymath}
  \varepsilon \ge (\nu^{-1}-1)\,{J_z\over 2}- 4\,E_b+ {\cal O}(t).
\end{displaymath}
In the limit we consider, this expression is a strictly decreasing
function of $\nu$; the optimal configuration of the stripe corresponds
to $\nu = 1$.    

\subsection{Implications}

The anisotropic $t$--$J$ model (\ref{t-J-Ham}) dominated by the Ising
term provides a good illustration to the strategy outlined in the
Introduction.  True, this simple model predicts insulating stripes
with $\nu=1$, contrary to experimental observations.  Nevertheless, it
has enabled us to find two possible phases of conducting stripes that
may arise in more realistic models: the 1D electron gas (the
``$\nu\to1$'' phase) suggested
previously\cite{Zachar-Kivelson,Kivelson-Fradkin} and the 1D holon
gas\cite{Pryadko-Tchernyshyov} (the ``$\nu\to0$'' phase).  To do so,
we have created a domain wall by fixing boundary conditions and doped
it to any given hole density $\nu$.  At this stage, the simplicity of
the model turns into a virtue: quantum numbers of elementary
excitations can be readily determined.\cite{Castro-Neto}

When macroscopic phase separation is absent, i.e., for $E_b>0$
($J_z<4\,V$ but $t,\,J_\perp\ll J_z$), a gas of holons is formed on a
weakly doped domain wall.  Holons are mobile kinks of the domain wall
with charge $Q=+1$ and no spin, $S_z=0$.  Compared to their
one-dimensional counterparts, domain-wall holons have an additional
flavor, $\rho=\pm1/2$ (isospin), which denotes the direction of the
associated transverse kink [Fig.~\ref{Fig-tJ-spinon}~(d)].  An
effective model describing the holon gas has been previously outlined
in Ref.~\CITE{Pryadko-Tchernyshyov}.

Even though it does not yield the observed stripe filling $\nu\approx
1/2$, the strongly anisotropic limit of the model~(\ref{t-J-Ham}) has
certain appeal: it is simple enough to permit controlled calculations.
In particular, spin waves are gapped and all associated dissipation
effects are suppressed.  In addition, holes are not allowed to leave
the domain wall and therefore cannot go around each other.  A domain
wall in this limit is a strictly one-dimensional object, yet its
transverse motion is fully accounted for.

\section{Holons on a domain wall: Hubbard model}
\label{Hubbard-numerics}

Could holons be generic to domain walls in an antiferromagnet or are
they just a curiousity of the $t$--$J$ model in the Ising limit?  To
answer this question, we have attempted to find similar excitations in
the Hubbard model.  Clearly, the problem is much more difficult
because there is no controlled approximation in this case, certainly
not in two spatial dimensions.

When the coupling strength $U$ is weak compared to the free-electron
bandwidth of $4t$, one can hope to find some guidance in mean-field
Hartree-Fock (HF) solutions.  This approach has been successful---to a
degree---in predicting the existence of stripes in the
cuprates\cite{early-HF}.  Numerical HF calculations show that, away
from half-filling, doped charges form stripes along directions (0,1)
or (1,1) with $\nu=1$ doped hole per lattice site along a stripe.
Such stripes are, indeed, AF domain walls.  Because they are filled
with holes to capacity, charged excitations are gapped and thus an
individual stripe is an insulator.\cite{Zaanen-note} (The same problem
arises in the $t$--$J$ problem in the large-$J_z$ limit considered
above.)

Holons described in Section \ref{Sec-t-J} are solitons with a
well-localized charge distribution.  To find analogous excitations in
the Hubbard model (at the HF level) let us recall the properties of
midgap states induced by domain walls.  It is well known that solitons
with anomalous quantum numbers arise in connection with fermion zero
modes induced by such topological defects\cite{Rajaraman} (see
Appendix \ref{app-0-modes}).  A uniform domain wall in 2
dimensions confines a midgap state only in one direction, across the
wall.  Solitons of finite extension---both across and along the
wall---can exist only if there is an inhomogeneity on the domain wall.
In this section we shall show that a bond-centered domain wall
with a wiggle may ``bind'' a holon, i.e., a soliton with quantized
charge $Q=1$ and spin $S_3 = 0$.

Although the solitons are static at the HF level, it is merely an
artifact of the mean-field approximation, in which the average spin
and charge densities are assumed to be time-independent.  Mean-field
configurations with solitons at different positions $x$ along the wall
should be viewed as degenerate minima of the action, i.e., as
classical solutions with a soliton at $x$.  Quantization of the
soliton restores broken translational symmetry: plane waves are
superpositions of states with a soliton at all possible sites $x$.  At
the semiclassical level, the energy of a soliton is given
by\cite{Rajaraman}
\begin{equation}
E = \sqrt{E_0^2 + p_x^2 v_{\rm 1D}^2} \approx E_0 + p_x^2 v_{\rm 1D}^2/2E_0,
\label{HF-holon-spectrum}
\end{equation}
where $E_0$ is the mean-field energy of the soliton and $v_{\rm 1D}$
is the 1D Fermi velocity calculated using the static HF wavefunctions.
As discussed below, the holon energy spectrum is similar to
that~(\ref{eq:holon-energy}) of the $t$--$J_z$ model, although both
the inverse mass and velocity are substantially reduced (in the
weak-coupling limit), reflecting the collective nature of the soliton.

Just as in the case of the anisotropic $t$--$J$ model, we are {\em
  not} dealing with the ground state of the model.  A stripe needs a
finite linear density of charge in order to be stable.  Because
collective excitations tend to be large at weak coupling, an
appreciable linear density of holons likely requires the coupling to
be strong, or else their overlap will completely destroy their
individual properties.  With only a Hartree-Fock approach at our
disposal, we cannot access the strong-coupling limit (although we have
tried to mimick it in the anisotropic $t$--$J$ problem).  Instead, we
maintain the domain wall by boundary conditions, and vary the linear
charge density along the stripe by changing the total number of holes
in the system.

The weak-doping expansion in the Hubbard model, however, has an
additional problem, which was {\em not\/} present in our analysis of
the $t$--$J_z$ model.  Namely, the undoped domain walls are always
unstable.  Indeed, the undoped system can be accurately described by a
Heisenberg-like model, and here the energy of a domain wall can be
continuously lowered by perturbing in the direction orthogonal to the
original magnetization vector.
In the static HF configurations presented below
[Figs.~\ref{Fig-spinon}--\ref{fig-1D-bands}], this untwisting
instability was suppressed by imposing a constraint of linear
polarization.  Nevertheless, these textures are the (particular)
solutions of the {\em full\/} set of HF equations.

The domain wall (antiphase stripe) is favored by the holes.  A
fully-doped stripe at $\nu=1$ is both locally and globally stable
already at the HF level.  For a {\em partially\/}-doped stripe, this
approximation (plus the constraint of linear polarization) gives
static localized holes.  We have found that the untwisting in the full
set of HF equations (arbitrary polarization) starts to develop on the
undoped portions of the stripe, compressing the remaining holes into
segments of a fully doped stripe ending with semi-vortices similar to
those discussed in Ref.~\CITE{Brazovski}.  We believe that this is an
artefact of the used weak-coupling approximation.  Namely, we expect
that at sufficiently strong coupling $U$ the effective attraction
between the holes (caused by the untwisting) will be compensated by
their increased mobility along the stripe.\cite{endnote:nn-repulsion}
In some sence this is similar to what happens in the anisotropic
$t$--$J$ model~(\ref{t-J-Ham}) in the region $-t\lesssim E_b<0$:
even though holes can gain some potential energy by sitting next to
each other, the associated loss of their kinetic energy prevents phase
separation along the stripe.

In the remainder of this section, we present numerical results
obtained in HF calculations.  Further analysis of the results in a
long-wavelength approximation is given in the next section.

\subsection{Hubbard model: numerical results}

We have solved self-consistently HF equations of the Hubbard model 
\[
-t \sum_{\Delta r}\psi({\bf r} + \Delta {\bf r}) 
- \frac{U}{2}[\langle\vec{\sigma}({\bf r})\rangle \cdot \vec{\sigma} 
- \langle\rho({\bf r})\rangle]\, \psi({\bf r}) = E\, \psi({\bf r}),
\] 
(the notation is explained in Sec.~\ref{Hubbard-analytics})
for collinear spin configurations,
\beq
\langle s_1({\bf r}) \rangle = \langle s_2({\bf r}) \rangle = 0,
\ \langle s_3({\bf r}) \rangle \neq 0, 
\label{collinear}
\eeq
at small and intermediate interaction strengths $U=2t\ldots4t$.  We
always started with two AF domains having opposite values of staggered
magnetization $m$.  On the lattice row separating the two domains,
staggered magnetization was initially disordered.  Such configurations
could thus later converge into site-centered, bond-centered, or
meandering stripes.

\subsubsection{Undoped stripe}  

At half-filling, the stripe always became
bond-centered, as in the anisotropic $t$--$J$ model.  We have
explicitly verified that a site-centered stripe always has a higher
energy in the absence of doped charges.  

\begin{figure}
\centering
\epsfxsize \columnwidth
\epsffile{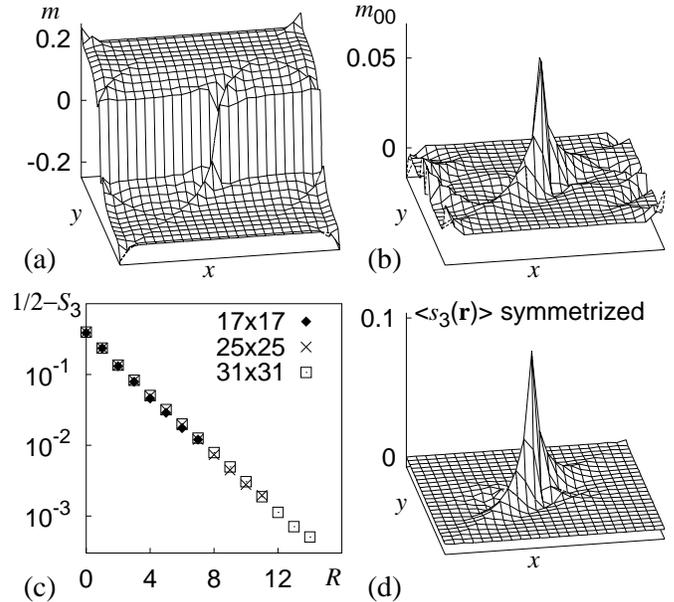}
\vskip 3mm
\caption{A spinon on a bond-centered wall with a wiggle.  Collinear HF
solution of the Hubbard model for $U=2t$ on a square lattice
$25\!\times\!25$.  (a) Staggered magnetization $m({\bf r})$.  (b)
Smoothed spin density $m_{00}({\bf r})$.  (c) The deviation of total
soliton spin $S_3(R)$, defined in \protect{Eq.~(\ref{S3R})}, from $1/2$ for
three different lattice sizes.  (d) Symmetrized spin density $\langle
s_3({\bf r}) + s_3({\bf -r})\rangle$/2.  }
\label{Fig-spinon}
\end{figure}

In a few cases, an initially disordered stripe converged to a state
with a higher energy, a bond-centered stripe with a defect where the
domain wall shifts one lattice spacing sideways
[Fig.~\ref{Fig-spinon}(a)].  Ripples in staggered magnetization
$(-1)^{x+y} \langle s_3({\bf r}) \rangle$ around the wiggle are an
interference effect between staggered spin, varying as $(-1)^{x+y}$,
and a smooth component of $s_3$.  By averaging $\langle s_3({\bf r})
\rangle$ over four neighboring sites,\cite{endnote:magnetization}  
\begin{eqnarray}
m_{00}({\bf r}) = &&[\langle s_3({\bf r})\rangle
+ \langle s_3({\bf r+\hat{x}})\rangle 
+ \langle s_3({\bf r+\hat{y}})\rangle
\nonumber \\
&&+ \langle s_3({\bf r+\hat{x}+\hat{y}})\rangle]/4,
\label{m00}
\end{eqnarray}
one can suppress the staggered component and uncover a spin soliton
residing at the wiggle [Fig.~\ref{Fig-spinon}(b)].  An even better
view of the soliton is afforded when spurious long-range spin-density
oscillations induced by the boundary are removed by plotting the
symmetrized spin density $\langle s_3({\bf r}) + s_3({\bf
-r})\rangle$/2 [Fig.~\ref{Fig-spinon}(d)].  Because particle density
remains equal to one everywhere, the soliton has zero charge (see
Appendix \ref{app-0-modes}).

Numerically, the soliton has a total spin $S_3=\pm1/2$:
\beq
S_3(R) = \sum_{|x|\leq R}\sum_{|y|\leq R}
\langle s_3({\bf r}) \rangle \to \pm1/2
\hskip 3mm \mbox{ as } R\to \infty.
\label{S3R}
\eeq
We have checked that the spin is well localized: $S_3(R)-S_3(\infty)$
vanishes exponentially with $R$ [Fig.~\ref{Fig-spinon}(c)].  This spin
soliton is an exact analogue of the spinon found in the $t$--$J$
problem [Fig.~\ref{Fig-spinon}(b)].  Here we also find spinons of two
flavors, those bound to transverse kinks and antikinks.

\subsubsection{Stripe with one doped hole}

\begin{figure}
\centering
\epsfxsize \columnwidth
\epsffile{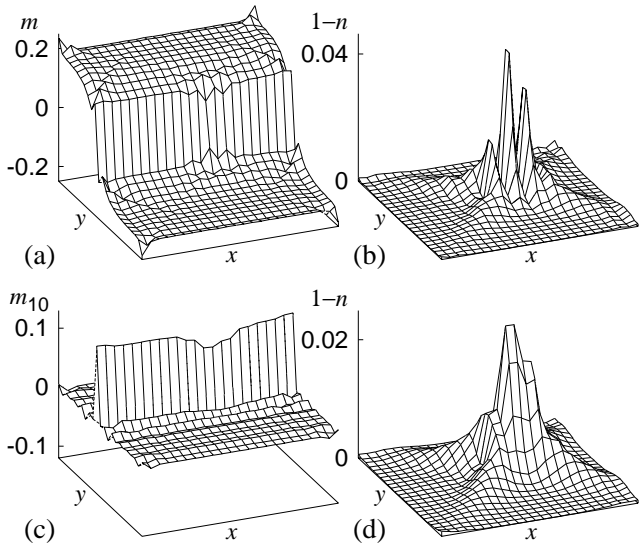}
\caption{A polaron on a straight bond-centered wall.  Collinear
  HF solution of the Hubbard model for $U=2t$ on a square lattice
  $25\!\times\!25$.  (a) Staggered magnetization $m({\bf r})$.  (b)
  Charge density $1-n({\bf r})$.  (c) $x$-staggered magnetization
  $m_{10}({\bf r})$. (d) Smoothed charge density.  }
\label{Fig-polaron}
\end{figure}

With one hole added, an initially disordered stripe typically
converged to one of the two bond-centered configurations.  A straight
bond-centered stripe would eventually contain a polaron
(Fig.~\ref{Fig-polaron}), a nontopological soliton with the quantum
numbers of a hole ($Q=1$, $S_3=\pm1/2$).  The presence of a nonzero
spin density is manifested in the typical ripples of the staggered
magnetization $(-1)^{x+y}\langle s_3({\bf r}) \rangle$---the result
of the interference between the staggered and smooth spin components.

\begin{figure}
\centering
\epsfxsize \columnwidth
\epsffile{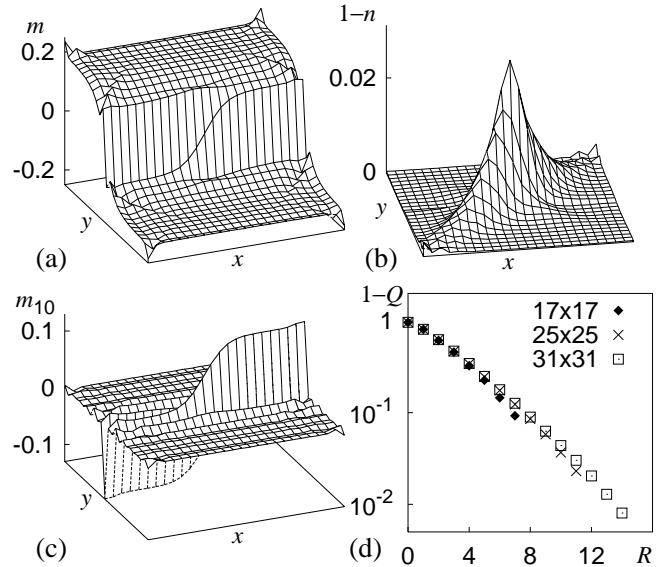}
\caption{A holon on a bond-centered wall with a wiggle.  Collinear
  HF solution of the Hubbard model for $U=2t$ on a square lattice
  $25\!\times\!25$.  (a) Staggered magnetization $m({\bf r})$.  (b)
  Charge density $1-n({\bf r})$.  (c) $x$-staggered magnetization
  $m_{10}({\bf r})$.  (d) The deviation of total charge $Q(R)$ around
  the soliton from 1 for three different lattice sizes. }
\label{Fig-holon}
\end{figure}

The other typical configuration was a bond-centered wall with a
transverse kink [Fig.~\ref{Fig-holon}], on which a charge $Q=+1$ is
localized.  The absence of interference fringes is a sign of zero spin
density.  Indeed, integrated spin $S_3(R)$ is numerically zero (of
order $10^{-13}$) for all values of $R$.  The absence of spin can also
be verified by plotting $m_{00}({\bf r})$, the smoothed spin density
(\ref{m00}).  Since $Q=+1$ and $S_3=0$, one immediately recognizes a
holon in this soliton.  As its counterpart in the anisotropic $t$--$J$
problem [Fig.~\ref{Fig-tJ-spinon}(d)], it binds to a transverse kink
or antikink of the domain wall.

Why does a wiggle on a domain wall change the nature of a doped charge
in such a dramatic way?  Essentially, a {\em bond}-centered domain
wall can be thought of as a 1D AF chain (roughly, two parallel spins
across the domain wall create an excess spin 1/2).  A straight domain
wall corresponds to a spin chain with perfect AF order.  If there is a
transverse kink on the domain wall, the staggered magnetization of the
effective chain changes sign at the transverse kink, i.e., staggered
magnetization itself has a kink.  In analogy with polyacetylene (as
discussed, e.g., by Berciu and John\cite{Berciu-John}), a doped charge
becomes either a polaron (no AF kink), or a holon (an AF kink is
present).  To illustrate this, we plot the $x$-staggered
magnetization\cite{endnote:magnetization} $m_{10}({\bf r})$
\begin{eqnarray*}
  &m_{10}({\bf r})
  = (-1)^x[\langle s_3({\bf r})\rangle 
  - \langle s_3({\bf r+\hat{x}})\rangle &
  \\
  & + \langle s_3({\bf r+\hat{y}})\rangle
  - \langle s_3({\bf r+\hat{x}+\hat{y}})\rangle]/4 &
\end{eqnarray*}
for a straight wall [Fig.~\ref{Fig-polaron}(c)] and for a wall
with a wiggle [Fig.~\ref{Fig-holon}(c)].  In both cases, the
$x$-staggered magnetization is confined to a narrow strip, which can
be identified with the effective chain of the $t$--$J$ problem.
Clearly, $m_{10}$ alters the sign at the center of a holon
[Fig.~\ref{Fig-holon}(c)] but is only slightly depressed around a
polaron [Fig.~\ref{Fig-polaron}(c)].  Later, we will substantiate
these qualitative arguments with an analysis of midgap states
(particularly, fermion zero modes).

\begin{figure}
\centering
\epsfxsize \columnwidth
\epsffile{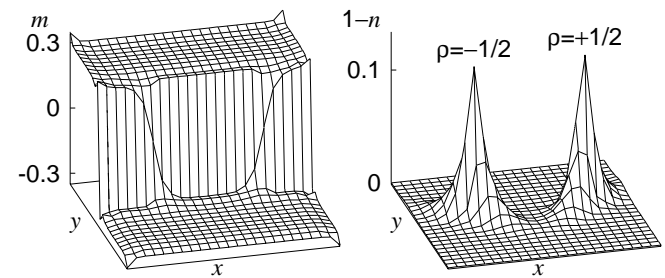}
\vskip 3mm
\caption{Staggered magnetization $m({\bf r})$ and hole density
$1-n({\bf r})$ in a HF calculation at $U=3t$,
$24\!\times\!24$ sites.  A bond-centered wall with 2 wiggles and 2
doped holes.}
\label{fig-2-holons}
\end{figure}

Although the question of stability of large solitons on a weakly doped
stripe is purely academic (see the discussion in the Introduction), we
have compared energies of isolated polarons and holons.  At weak
coupling, $U\lesssim 2.5t$, polarons have a slightly lower energy.
Two or more polarons preferred to bind into spinless bipolarons.  For
$U\geq 3t$, holons had a lower energy.  Moreover, two well-separated
holons (Fig.~\ref{fig-2-holons}) had a smaller energy than a bipolaron
for $U\geq3t$.  The trend is clearly to favor holons as the coupling
gets stronger.

\section{Hubbard model: a Hartree-Fock analysis}
\label{Hubbard-analytics}

To interpret the numerical results discussed in the previous section,
we have conducted a thorough analysis of electron states induced in
the middle of the Hubbard gap by an AF domain wall.  Midgap states of
a straight domain wall have been previously explained in great detail
by Schulz.\cite{Schulz} Studying a domain wall with a wiggle on a
lattice is a rather challenging task, therefore we first derive a
long-wavelength approximation that could capture the essential physics
(e.g., the difference between bond-centered and site-centered walls).
We find that midgap states of an undoped domain wall resemble a
half-filled 1D chain with the Fermi momentum $k_F=\pi/2$.  The
relevant low-energy states on a domain wall are composed of Bloch
states with momenta near $(\pm\pi/2, \pm\pi/2)$.  This leads to a
theory of 8-component fermions (2 spin components $\times$ 4 Fermi
points).  Finally, we relate holons observed numerically to fermion
zero modes induced by a wiggle, and trace the origin of holon
transverse flavor to the doubling of fermion components (8 instead of
the usual 4 in the 1D case).

\subsection{Mean-field equations}

The HF equations for the Hubbard model are
\beq
-t \sum_{\Delta r}\psi({\bf r} + \Delta {\bf r}) 
- \frac{U}{2}[\langle\vec{\sigma}({\bf r})\rangle \cdot \vec{\sigma} 
- \langle\rho({\bf r})\rangle]\, \psi({\bf r}) = E\, \psi({\bf r}),
\label{HF_full}
\eeq
where $\psi({\bf r})$ is a 2-component spinor wavefunction,
$\vec{\sigma}$ is the triplet of Pauli matrices and the sum is over
vectors $\Delta{\bf r}$ pointing to the four adjacent sites.  The
expectation values of spin $\langle\vec{\sigma}({\bf r})\rangle$ and
density $\langle\rho({\bf r})\rangle$ should be calculated
self-consistently.  As discussed in the Introduction, we are
specifically looking for collinear\cite{Schulz,endnote:collinear}
solutions, therefore we set $\langle\sigma_1\rangle =
\langle\sigma_2\rangle = 0$.  It is customary to rotate the spin axes
on one of the sublattices through $\pi$, which we do as follows:
\beq
\psi({\bf r}) \to 
\sigma_1^{x+y}\psi({\bf r}).
\label{staggered-rotation}
\eeq
In the new basis, the HF equation reads
\beq
-\sigma_1 t\sum_{\Delta {\bf r}} \psi({\bf r} + \Delta {\bf r}) 
- U[\sigma_3 m({\bf r}) - n({\bf r})/2]\, \psi({\bf r}) = E\, \psi({\bf r}),
\label{HF-3}
\eeq
where mean-field parameters $m({\bf r})$ and $n({\bf r})$ are the
staggered spin and the charge density.

To simplify further analysis, we neglect density
fluctuations\cite{neglect-rho} and set $n({\bf r}) = n$ in
Eq.~(\ref{HF-3}), also shifting $E \to E+Un/2$.  The resulting HF
equation,
\beq 
-\sigma_1 t\sum_{\Delta {\bf r}} \psi({\bf r} + \Delta {\bf r}) - \sigma_3\, U
m({\bf r})\, \psi({\bf r}) = E\, \psi({\bf r}),
\label{HF-3-symm}
\eeq
acquires a charge conjugation symmetry
\beq
\psi({\bf r}) \to \sigma_2 \psi^*({\bf r}).
\label{C}
\eeq
In this particle--hole symmetric form, the mean-field equations
resemble those in the theory of polyacetylene.\cite{SSH,Maki} The
discrete symmetry (\ref{C}) has important implications, among them
the possibility of spin-charge separation.

When staggered magnetization is uniform, the HF Hamiltonian
(\ref{HF-3-symm}) can be readily diagonalized using the momentum
basis:
\[
E_{\bf k} = \pm \sqrt{\epsilon_{\bf k}^2 + U^2m^2},
\]
where $\epsilon_{\bf k} = -2t(\cos{k_x} + \cos{k_y})$.  The
one-electron energy spectrum has a gap $2\Delta_0 = 2U|m|$, which
makes the system an insulator at half-filling.  Note that, thanks to
the spin axis rotation (\ref{staggered-rotation}), there is no
doubling of the unit cell and the Brillouin zone is therefore not
folded.  

\subsection{Straight domain wall along $x$}

A domain wall implies a change of sign for staggered magnetization
$m({\bf r})$ along a line on the lattice.  Because the ``local gap''
$U|m({\bf r})|$ is reduced on this boundary, one expects to find
electron states inside the Hubbard gap $\Delta_0$.  In the case of an
isolated straight uniform domain wall, such states are localized
across the wall and extended along it.  In view of the charge
conjugation symmetry (\ref{C}), Eq.~(\ref{HF-3-symm}) has an even
number of midgap states for every momentum $-\pi < k_x \leq \pi$,
normally two.\cite{Schulz}

\begin{figure}
\noindent\leavevmode\epsfxsize\columnwidth
\epsffile{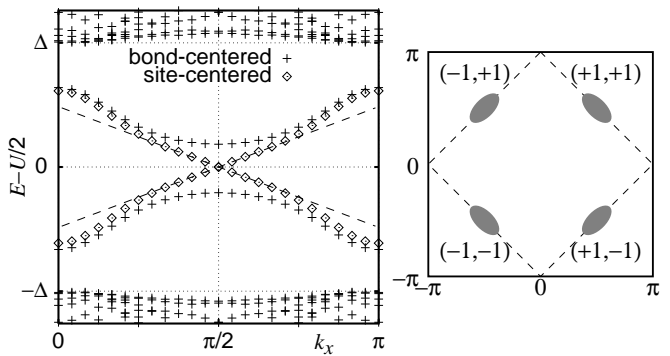}
\vskip 2mm
\caption{Left: midgap one-particle spectrum $E(k_x)$ of an AF with a
  straight domain wall.  Self-consistent solution of HF
  equations.  $U=2.5t$, $48\!\times\!27$ sites.  Dashed line:
  low-energy approximation, Eq.~(\protect\ref{H_eff_scw}).  States
  outside the gap are not shown for the site-centered stripe.  Right: 4
  Fermi patches in the Brillouin zone, Eq.~(\protect\ref{trispinor}),
  shown with eigenvalues $(\tau^x_3,\tau^y_3)$.  Dashed line: Fermi
  surface of the noninteracting system.  }
\label{fig-1D-bands}
\end{figure}

Depending on the symmetry of the domain wall---it can be {\em site}
or {\em bond}-centered---the two midgap bands intersect at
$k_x=\pm\pi/2$ or are separated by a smaller gap $2\Delta_1$,
respectively (Fig.~\ref{fig-1D-bands}).  Qualitatively, this can be
understood by picturing electrons in the domain-wall bands as a 1D
electron gas in an external magnetic field $Um(x,y) = B(y)\cos(\pi
x)$, which is either antisymmetric (site-centered), or symmetric
(bond-centered) in $y$.  In the former case, $B(y)$ averaged across
the wall vanishes and we deal with effectively free 1D electrons.  In
the latter case, the electrons feel a nonzero magnetic field staggered
along $x$, which induces the smaller gap $2\Delta_1$ near
$k_x=\pm\pi/2$.  A more comprehensive discussion can be found in
Appendix \ref{straight-wall-app}.

An undoped antiferromagnet is particle--hole symmetric, therefore only
a half of the midgap states are filled.  Since the total number of
midgap states on a wall of length $L$ is $2L$, the maximum linear
density of holes or electrons that can be doped is $\nu=1$.  Once all
the midgap states are filled, the system is again an insulator.

Stability of the ``magic'' filling $\nu=1$ can be explained by the
following qualitative argument.  There is only one energy scale
$\Delta_0$ in the limit of weak coupling. The preferred filling is
determined by minimizing the total energy cost---including the energy
of domain walls---per doped particle.  When $\nu<1$, holes are doped
into a midgap band, where one-particle energies are much less than the
Hubbard gap $\Delta_0$---at least in the weak-coupling limit.
Therefore, the main part of the energy cost comes from creating a
domain wall, which should be of order $\Delta_0$ per unit length:
\[
\varepsilon(\nu) \equiv \frac{E}{\nu L} \approx \frac{\alpha \Delta_0}{\nu},
\hskip 5mm
\nu < 1,
\]
$\alpha$ is of order 1 (in 1D, $\alpha = 2/\pi$).  Doping beyond
$\nu=1$ puts holes into the lower Hubbard band, separated by the gap
$\Delta_0$, at a cost
\[
\varepsilon(\nu) 
= \frac{\alpha \Delta_0 + (\nu-1)\Delta_0}{\nu},
\hskip 5mm
\nu > 1.
\]
At $\nu=1$, the energy per doped hole has a cusp, where 
the derivative $\varepsilon'(\nu)$ jumps from $-\alpha\Delta_0$ 
to $(1-\alpha)\Delta_0$.  The cusp is actually a minimum: 
since stripes with $\nu=1$ are known to exist, they must have a lower
energy per doped hole than the uniform AF state, $\varepsilon(1) < \Delta_0$,
i.e., $\alpha<1$.  

The minimum of $\varepsilon(\nu)$ may shift to a lower filling if there
are two energy scales for carriers on a stripe or if a fairly large
gap (comparable to $\Delta_0)$ opens up in the 1D band at some value
of $\nu<1$.  An example of the former scenario, described by Nayak and
Wilczek, gives a smooth minimum at an incommensurate filling
determined by the ratio of the two energy scales; this yields
conducting stripes.  In the latter case, a large gap can possibly be
the result of a commensurate filling $\nu_0 < 1$, producing a cusp in
$\varepsilon(\nu)$; such stripes will likely be insulating.  Neither
argument provides a convincing explanation for experimentally observed
{\em conducting\/} stripes with the ``magic'' filling fraction
$\nu=1/2$.

\subsection{Continuum formulation}
\label{subsection-continuum}

In the weak-coupling limit, $U\ll 4t$, the characteristic distances
are large, and a continuum approximation of some sort should provide a
sufficiently accurate description of the system.  The approximation
must be intelligent enough to tell apart, say, bond-centered and
site-centered domain walls, which, as we have seen, have quite
different one-particle spectra.  The difference comes from a change in
the symmetry of the domain wall, and it should be possible to
describe within a continuum theory.

To construct an effective continuum approximation for describing a
weakly-deformed domain wall, we first have to identify the relevant
modes.  In a weakly-coupled Hubbard model near half-filling, all
low-lying excitations are concentrated near the Fermi-lines $|k_x\pm
k_y|=\pi$.  A domain wall induces a 1D midgap electron band, which is
half-filled ($k_F=\pi/2$) if the domain wall is not doped.  The
relevant modes for describing a weakly-deformed domain wall should be
located near the intersection of the two Fermi-lines, which gives four
``Fermi points'' ${\bf k} = (\pm\pi/2, \pm\pi/2)$.  Together with the
spin index, this implies that the continuum description should be
formulated in terms of $8$-component Fermion wavefunctions.  An
alternative, more quantitative way to reach the same conclusion is
presented in Appendix~\ref{straight-wall-app}, where a straight domain
wall on the lattice is analyzed.

We write an electron wavefunction as a sum of four terms with smoothly
varying amplitudes:
\begin{equation}
\psi_s({\bf r}) \approx 
\sum_{\alpha=\pm1} \sum_{\beta=\pm1}\ 
\psi_{\alpha\beta s}({\bf r})e^{i\pi (\alpha x + \beta y)/2}.
\label{trispinor}
\end{equation}
[As there is no folding of the Brillouin zone in our formalism, points
$(\pi/2,\pi/2)$ and $(-\pi/2,-\pi/2)$ are not equivalent.]  In
Eq.~(\ref{trispinor}), we have added two more indices, $\alpha={\rm
sgn}\,k_x$ and $\beta={\rm sgn}\,k_y$ to the staggered spin index $s$
[Eq.~(\ref{staggered-rotation})] .  Only those Fourier components of
magnetization which connect the four Fermi patches are preserved:
\[
\langle s_3({\bf r})\rangle \approx 
{ \sum_{\alpha=0}^{1} \sum_{\beta=0}^{1}}\ 
m_{\alpha\beta}({\bf r})e^{i\pi (\alpha x + \beta y)}.
\]
Each index has its own set of Pauli matrices, $\{\tau^x_i\}$,
$\{\tau^y_i\}$, and $\{\sigma_i\}$ for $\alpha$, $\beta$, and $s$,
respectively ($i=1,2,3$).  Any two operators from different sets 
commute with each other because they act on different indices.  
Some of these operators have a transparent physical meaning:
\begin{eqnarray}
{\bf k} \approx \frac{\pi}{2}(\tau^x_3, \tau^y_3),
\hskip 5mm
(-1)^x = \tau^x_1,
\hskip 5mm
(-1)^y = \tau^y_1,
\label{phys-meaning}
\\
s_1 = \frac{\sigma_1}{2}, 
\hskip 5mm
s_2 = \frac{\sigma_2 e^{i{\bf Q\cdot r}}}{2}, 
\hskip 5mm
s_3 = \frac{\sigma_3 e^{i{\bf Q\cdot r}}}{2}.
\end{eqnarray}
Using this notation we write the continuum approximation of the 
HF Hamiltonian (\ref{HF-3-symm}) as
\begin{eqnarray}
  H_{\rm HF} = 
-2ita\,\sigma_1(\tau^x_3\partial_x + \tau^y_3\partial_y)
- U\sigma_3 m({\bf r}), 
\label{HF_continuum_2}
\\
m \equiv m_{11} + m_{01}\tau^x_1 
+ m_{10}\tau^y_1 + m_{00}\tau^x_1\tau^y_1,
\nonumber
\end{eqnarray}
where $a$ is the lattice constant.  
Only $m_{11}({\bf r})$, staggered magnetization proper, exists in
the bulk of the antiferromagnet inducing the Hubbard gap $\Delta =
U|m_{11}(\infty)|$.  The energy spectrum near the 4 points $(\pm\pi/2,
\pm\pi/2)$ has the form
\[
E^2 = 4t^2a^2(p_x \pm p_y)^2 + U^2m_{11}^2.
\]
The other three components of $m$ (e.g., the average spin density
$m_{00}$) can be induced around defects only.  As we will show, states
localized on defects are particularly sensitive to these components.

\subsection{Midgap spectrum of a straight wall}

To warm up, let us derive the midgap spectrum of a straight domain
wall along the $x$ direction.  Translational invariance requires that
$m_{01}=m_{00} = 0$.  From lattice solutions (Appendix
\ref{straight-wall-app}) we know that the wall fermions have a gapless
(gapped) energy spectrum for a site-centered (bond-centered) domain
wall.  The absence of a gap can be demonstrated by finding a fermion
mode with zero energy at $p_x\equiv -i\partial_x=0$ (i.e.,
$k_x=\pm\pi/2$).  The Schr\"odinger equation (\ref{HF_continuum_2}) 
for $E=0$ reads
\begin{equation}
\sigma_2 \frac{d\psi}{dy} = \frac{U}{2ta} 
[\tau^y_3 m_{11}(y) + i\tau^y_2 m_{10}(y)] \psi(y). 
\label{eq_0_modes}
\end{equation} 
If we neglect $m_{10}$ at first, solutions of Eq.~(\ref{eq_0_modes})
are eigenstates of $\sigma_2$, $\tau^y_3$ and, e.g., $\tau^x_3$ (then
each zero mode comes from a single Fermi patch).
Eq.~(\ref{eq_0_modes}) reduces to 8 uncoupled scalar equations, giving
a total of 8 linearly independent solutions.  As
usual,\cite{Rajaraman} only half of them [those with eigenvalues
$\sigma_2\tau^y_3 m_{11}(+\infty)<0$] are localized on the wall, so
that there are 4 zero modes.

Remarkably, in addition to the usual twofold spin degeneracy, there is
another spin-like degree of freedom, which will prove to be the
transverse flavor.  The origin of isospin (at weak coupling) is thus
exposed: compared to a 1D chain, there are twice as many ``Fermi
points'' on a straight domain wall in 2D---see Fig.~\ref{fig-1D-bands},
right.  

The difference between midgap spectra of site-centered and
bond-centered walls arises already in the first order in $m_{10}$.
Eq.~(\ref{eq_0_modes}) has four zero modes if $m_{10}$ is an odd
function of $y$, i.e., for a site-centered wall.\cite{Bloch}
Otherwise, the middle band is split by a small gap
(Fig.~\ref{fig-1D-bands}, left).

\subsubsection{Site-centered domain wall}

In the presence of a nonvanishing $m_{10}$, the spectrum $E(p_x)$ of a
straight site-centered stripe can be determined approximately by
starting with $p_x=0$ [Eq.~(\ref{eq_0_modes})] and treating the term
containing $\partial_x$ in the Hamiltonian~(\ref{HF_continuum_2})
perturbatively.  In the limit $p_x\to 0$, states outside the main gap
can be neglected, which reduces the Hilbert space to the four zero
modes [Eq.~(\ref{eq_0_modes})].  By using the degenerate perturbation
theory, we find a Dirac spectrum (dashed lines in
Fig.~\ref{fig-1D-bands}, left): as $p_x\to 0,$
\begin{equation}
E(p_x) \sim \pm v p_x, 
\hskip 3mm
v=2ta \langle\tau^y_1\rangle.
\label{H_eff_scw}
\end{equation}
This compares well with a similar result (\ref{t1D}) obtained on a
lattice.

\subsubsection{Bond-centered domain wall}

Alternatively, one can explore the limit of a small $x$-staggered
magnetization, $m_{10} \ll m_{11}(\infty)$.  In that case, by starting
with $m_{10}=0$, one finds 4 degenerate zero modes for any $p_x$.  A
nonzero $m_{10}$ induces a splitting of the zero modes.  To lowest
order in $m_{10}$, the energy at $p_x=0$ is
\[
E(0) = \pm\Delta_1 = \pm U \langle m_{1} \rangle
\equiv \pm U\int m_{10}(y) \psi^\dagger(y)\psi(y) dy.
\]
As claimed, the gap is proportional to the $x$-staggered
magnetization felt by an electron on the domain wall.  

\subsection{Zero modes at a wiggle}

One way to prove that a charged soliton at a wiggle indeed has zero
spin is to show that the HF equations contain a doubly degenerate
fermion zero mode.  An elementary discussion of the connection between
zero modes and separation of spin and charge is given in Appendix
\ref{app-0-modes}.  Because the problem is essentially two-dimensional
(the domain wall is curved), it is much harder than its 1D analogs.
In 1D, symmetry arguments are generally sufficient to prove the
existence of zero modes in 1D---even on a lattice!  (See Appendix
\ref{straight-wall-app}.)  In contrast, we have not been able to find
such a general proof for the 2D problem of a wall with a
wiggle---neither on the lattice, nor in the long-wavelength
approximation.

Instead, we offer a somewhat hand-waving argument in favor of zero
modes in this case.  Lack of rigor is compensated by an insight 
into the origin of the transverse flavor: it turns out that holons
residing on transverse kinks and antikinks come from different 
points of the Brillouin zone, i.e., they are made of completely 
different stuff.  

\begin{figure}
\centering
\leavevmode
\epsfxsize 0.9\columnwidth
\epsffile{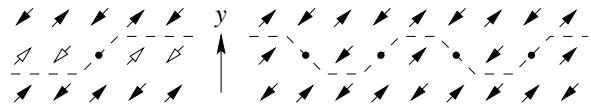}

\vskip 5mm
\caption{Left: Bond-centered stripe with a wiggle 
as a superposition of a site-centered domain wall (black arrows) and a 
1D AF chain with a kink (open arrows).  
Right: Holons with alternating isospins form a cite-centered stripe. 
}
\label{fig_site_to_bond}
\end{figure}

As illustrated in Fig.~\ref{fig_site_to_bond}, magnetization on a {\em
bond\/}-centered wall with a wiggle can be obtained by superimposing
$m({\bf r})$ of a straight {\em site\/}-centered wall and that of a
spin chain with a kink in $x$-staggered magnetization.  Away from the
wiggle, $m_{00}({\bf r})= m_{01}({\bf r}) = 0$.  To simplify the
discussion, we will neglect these components altogether
(but this is nonessential and can be remedied).  
Decompose $m({\bf r})$ into an $x$-independent
part and the rest:
\begin{eqnarray}
m({\bf r}) &=& m^{(0)}(y) + m^{(1)}({\bf r}),
\nonumber\\
m^{(0)}(-y) = -m^{(0)}(y),
&& 
m^{(1)}(\pm\infty,-y)=m^{(1)}(\pm\infty,y).
\nonumber
\end{eqnarray}
The Hamiltonian (\ref{HF_continuum_2}) can now be split in two parts:
\begin{eqnarray}
H^{(0)}_{\rm HF} = -2ita\, \sigma_1 \tau^y_3 \partial_y 
-U\sigma_3 [m^{(0)}_{11}(y) + m^{(0)}_{10}(y) \tau^y_1],
\label{H0}\\
H^{(1)}_{\rm HF} = -2ita\, \sigma_1 \tau^x_3 \partial_x
-U\sigma_3 [m^{(1)}_{11}({\bf r}) + m^{(1)}_{10}({\bf r}) \tau^y_1]. 
\label{H1}
\end{eqnarray}
As shown above, the ``transverse part'' (\ref{H0}) has 4 zero modes
for each $p_x$.  Within this Hilbert space, $H^{(1)}_{\rm HF}$ describes 
right and left-moving fermions with spin, which see a staggered magnetization
$$\langle m_{1}(x) \rangle = 
\int\!dy\ u^\dagger(y)
[m_{10}({\bf r}) + m_{11}({\bf r})\tau^y_1]u(y),$$
where $u(y)$ is a zero mode (\ref{eq_0_modes}) of Eq.~(\ref{H0}).  The
midgap fermion band acquires a gap of its own, 
$\Delta_1 = U|\langle m_{1}(\infty)\rangle| < \Delta$,
with two zero modes (one for each spin) inside this smaller gap.
``Longitudinal'' wavefunctions of the two zero modes satisfy the equation 
\begin{equation}
\sigma_2 \frac{d\psi(x)}{dx} 
= \frac{U}{2ta\langle\tau^y_1\rangle}\,\tau^x_3 
\langle m_{1}(x) \rangle\, \psi(x).
\label{eq_0_modes_wall}
\end{equation}

The existence of two holon flavors can now be deduced from 
Eqns.~(\ref{eq_0_modes}) and (\ref{eq_0_modes_wall}).  The zero modes
have a finite norm only if
$$\sigma_2\tau^y_3 m^{(0)}_{11}(+\infty) < 0, 
\hskip 3mm
\sigma_2\tau^x_3 
\langle m_{1}(+\infty) \rangle / \langle\tau^y_1\rangle < 0.$$
It follows then that the product of eigenvalues 
\begin{equation}
\tau^x_3\tau^y_3 = 
{\rm sgn} [ m^{(0)}(y=+\infty) \, \langle m_1(x=+\infty)\rangle \,
\langle\tau^y_1\rangle ]
\label{signs}
\end{equation}
can be identified with the holon isospin $2\rho$.  This can be seen by
extrapolating Eq.~(\ref{signs}) to larger values of $U$, which reduces
the size of holons.  We have $\langle\tau^y_1\rangle =
\langle(-1)^y\rangle = (-1)^{y_0}$, where $y_0$ is the row number of
the chain in Fig.~\ref{fig_site_to_bond}.  According to
Eq.~(\ref{signs}), if $\tau^x_3\tau^y_3 = +1$, spins on the chain and
to the right (left) of the wiggle are an extension of the upper
(lower) AF domain, as for the $\rho=+1/2$ wiggle in
Fig.~\ref{fig_site_to_bond}.  Thus, $\rho = \tau^x_3\tau^y_3/2$.  This
identification is consistent with numerical HF solutions
(Fig.~\ref{Fig-holon}), where $\tau^x_3\tau^y_3 = {\rm sgn} k_x\, {\rm
sgn} k_y$ can be inferred from the orientation of a holon---the
perfect nesting of the Fermi surface makes holon wavefunctions 
cigar-shaped and oriented along a lattice diagonal.

\section{Conclusions}

We have attempted to infer a set of plausible quantum numbers of
low-lying excitations on a partially doped domain wall in a
strongly-correlated antiferromagnet by analyzing artificially-created
domain walls in simpler systems.  Specifically, we have studied
quantum numbers of well-separated holes doped into domain walls in the
$t$--$J$ model with Ising anisotropy\cite{Kivelson-tJz} and in the
Hubbard model (in a Hartree-Fock approximation).  In addition to a
usual 1D electron gas,\cite{EK} we have identified a new potential
candidate: the 1D gas of holons.  In this phase, which we have found
at sufficiently small linear hole density $\nu$ in both models, charge
carriers (holons) have spin $S_3=0$ and charge $Q=+1$.  Each holon
resides on a transverse kink of the domain wall, which leads to a
strong interplay between charges and transverse fluctuations of a
stripe.

We find it very encouraging that the charge carriers with identical
quantum numbers result from two vastly different calculations.  In the
strongly coupled $t$--$J_z$ model, the holons are small and
immediately evident [Fig.~\ref{Fig-tJ-spinon}(d)].  In the
weakly-coupled Hubbard model, they are large and represent a
collective effect (fermion zero modes).  This indicates that a
universality of some sort is at play, and, therefore, that the same
``$\nu\to0$'' phase could result from more authentic models.  Whether
or not this phase is relevant for the cuprate stripes, which have
$\nu\approx1/2$, remains an open question.  In future, we intend to
extend this work to intermediate linear hole densities.

\section*{Acknowledgments}

The authors thank C. Chamon, M.~M. Fogler, S.~A.  Kivelson,
F. Wilczek, and J. Zaanen for valuable discussions.
Part of this work has been done at the Aspen Center for Physics.
Financial support from DOE Grant DE-FG02-90ER40542 is gratefully
acknowledged.

\appendix

\section{More on the $t$-$J_z$ model}
\label{app:lower-bound}

In this appendix we give some estimates of the energies of domain
walls in the strongly anisotropic $t$-$J$ model~(\ref{t-J-Ham}).  In
particular, we show that in this limit doped holes can lower their
energy by forming (fully-packed) domain walls.  This possibility has
not been considered in Ref.~\CITE{Kivelson-tJz}.  We also discuss 
stability of the dilute holon gas considered in Sec.~\ref{Sec-t-J}.

In the limit of infinite $J_z$, only fully doped domain walls, similar
to those considered by Osman {\em et al.},\cite{insulating-stripes}
have a finite energy (all frustrated bonds are covered with holes).  A
domain wall must therefore maintain continuity, i.e., adjacent holes
must be nearest or next-nearest neighbors.  This implies that a
domain wall horizontal on average, can change its height $y(x)$ by at
most one unit at a time.  Such a domain wall can be fully described
using a one-dimensional language, namely by specifying differences in
the heights of neighboring holes:
\[
y(x+1)-y(x) = -1, 0, \mbox{ or } 1.  
\]
The system is therefore equivalent to a spin-$1$
chain,\cite{insulating-stripes} with the Hamiltonian
\begin{equation}
  \label{eq:spin-one-chain}
  H_{\rm eff}=\sum_n \left[-3\,E_b-E_b\,\left(S_n^z\right)^2
  +{t\over2}\, \left(S_n^+\,S^-_{n+1}+{\rm h.c.}%
  \right)\right], 
\end{equation}
where $n=x+1/2$ and $E_b=V-J_z/4$ is the energy of an AF bond.
The first term in Eq.~(\ref{eq:spin-one-chain}) represents the energy
of a straight segment of a wall (three broken bonds per hole), the
second term counts the number of additional broken bonds due to kinks,
while the last term describes the transverse hops %
of holes.

The properties of the effective spin
Hamiltonian~(\ref{eq:spin-one-chain}) and its generalizations have
been extensively studied.\cite{insulating-stripes,spin-chain-papers}
When $E_b\gg t$, the AF ordering $S_n^z=(-1)^n$ (a zigzag wall) is
favored.  For $E_b$ large and negative the ground state corresponds to
$S_n^z=0$ (a flat domain wall).  At $|E_b|\lesssim t$, the system
enters an intermediate critical phase, in which the density of domain
wall kinks varies continuously.

Up to terms of higher order in $t/|E_b|$, the energies (per hole)
of the two ordered phases are
\begin{equation}
  \label{eq:ordered-spin-one-energies}
  \varepsilon_{{\rm flat}}=-3\,E_b-t^2/|E_b|,\quad
  \varepsilon_{{\rm zigzag}}=-4\,E_b-t^2/E_b.
\end{equation}
For $E_b<0$ (two holes attract), phase separation in the bulk affords
a lower energy, $-2E_b < \varepsilon_{\rm flat}$, thus hindering
natural formation of stripes.  For a strongly repulsive bond, $E_b\gg
|t|$, stripes win over a lump of immobilized holes: transverse
fluctuations of a zigzag stripe lower its kinetic energy by an amount
of order $t$ per hole [cf. Eq.~(\ref{eq:bulk-localized-hole})].  This
possibility has not been considered in Ref.~\CITE{Kivelson-tJz}.

Let us now consider a domain wall created artificially ({\em e.g.}, by
boundary conditions at infinity) in order to study partially doped
domain walls.  In the absence of holes, such a wall is {\em
bond}-centered and straight to minimize the number of broken bonds
[see Fig.~\ref{Fig-tJ-spinon}a].  The corresponding energy cost per
unit length is $J_z/2$.

When holes are added to an antiferromagnet with a domain wall, they
will necessarily bind to it (each hole on the domain wall reduces the
number of frustrated spins; the energy gain is $\sim J_z/2$ per hole.)
As discussed in Sec.~\ref{Sec-t-J}, a single doped hole acquires
mobility by riding a kink [See Fig.~\ref{Fig-tJ-spinon}d].  The
corresponding energy is given by Eq.~(\ref{eq:holon-energy}).
Assuming that holons are well separated, the energy per added charge
is
\begin{equation}
  \label{eq:holon-gas-energy}
  \varepsilon_{\rm holon\;gas}=-4E_b-2t+{\cal
  O}(t^2/J_z,\,\nu \,t).
\end{equation}
Here we have not included the energy cost of creating a domain wall.  

The assumption of large separation may be violated even at very small
linear densities $\nu\ll1$ if there is an attractive interaction
between holes.  For example, for $E_b\lesssim -t$, the
energy~(\ref{eq:ordered-spin-one-energies}) of the fully-packed flat
phase can be smaller than that of the holon gas,
(\ref{eq:holon-gas-energy}).  The holes on the stripe will separate 
into a dense phase ($\nu=1$) leaving a portion of the domain wall 
comletely undoped ($\nu=0$).
  
To find a strict upper bound $E_b^{\rm min}$, such that for
$E_b<E_b^{\rm min}$ the phase separation definitely happens, we can
use a variational estimate for the ground state energy of the
Hamiltonian~(\ref{eq:spin-one-chain}).  The simplest estimate
corresponds to all $S_n^z=0$, which immediately gives
$\varepsilon<-3E_b$.  This is smaller then the energy of a dilute
holon gas (\ref{eq:holon-gas-energy}) for $E_b< E_b^{\rm min}=-2t$.
This estimate of the phase separation boundary $E_b^{\rm min}$ can be
easily improved (increased) by using more sophisticated 
variational wavefunctions.

On the other hand, phase separation of this sort is {\em not\/}
expected for $E_b>0$, when holes have uniformly {\em
repulsive\/} interactions.  This statement can be made more formal by
evaluating a strict {\em lower\/} bound on the energy of any dense
hole phase described by the Hamiltonian~(\ref{eq:spin-one-chain}).
Estimating each term in the Hamiltonian independently, we have, for
$E_b>0$,
$$
H_{\rm eff}>\sum_n\left(-4E_b-2|t|\right)= -N_h\,\varepsilon_{\rm
  holon\;gas}.
$$
The inequality is strict because the terms in the original
Hamiltonian do not commute.  It implies that
a dilute holon gas is stable to phase separation to a completely doped
region and a region of an undoped stripe, as expected on physical
grounds for a repulsive interaction.

\section{Zero modes and separation of spin and charge}
\label{app-0-modes}

We retrace the relation between fermion zero modes and separation of
spin and charge.\cite{SSH,Maki} For completeness, we will use the
lattice version of the HF equations (\ref{HF-3-symm}), \beq -\sigma_1
t\sum_{\Delta {\bf r}} \psi({\bf r + \Delta {\bf r}}) - \sigma_3\, U
m({\bf r})\, \psi({\bf r}) = E\, \psi({\bf r}).
\label{the-HF}
\eeq

\subsection{Symmetries}

So long as we deal with collinear AF configurations, one component of
spin---$(\sigma_3/2) \exp{(i\,{\bf Q\!\cdot\!r}})$ in our
notation---is a conserved quantity.  The transformation
\beq
\psi({\bf r}) \to \sigma_3\, e^{i\,{\bf Q \cdot r}} \psi({\bf r})
\label{symm-sigma3}
\eeq
is a symmetry of the mean-field Hamiltonian.  Here ${\bf Q} =
(\pi,\pi)$.

The unitary part of charge conjugation (\ref{C}),
\beq
\psi({\bf r}) \to \sigma_2 \psi({\bf r}),
\label{symm-sigma2}
\eeq
alters the sign of $E$ and as such is not a symmetry of the 
mean-field equations.  Rather, it can be referred to as a 
symmetry of zero modes.  

For the sake of convenience, we also want the system to be reasonably
symmetric with respect to some sort of a parity transformation ${\bf
r} \to -{\bf r}$.  If, however, the system has a domain wall, staggered
magnetization will be antisymmetric under parity.  To make it a symmetry,
we combine parity with a spin flip:
\begin{equation}
\psi({\bf r}) \to \sigma_1 P\, \psi({\bf r}) \equiv \sigma_1\,\psi({-\bf r}).
\label{CP}
\end{equation}
This ``combined parity'' is a symmetry of HF equations,
provided that $m(-{\bf r}) = -m({\bf r})$.

\subsection{Fermion zero modes}

Under some circumstances, Eq.~(\ref{the-HF}) has solutions with zero
energy.  Such solutions are normally localized on a topological
defect, e.g., on a domain wall. Note that a straight domain wall
confines a fermion zero mode in one direction only---across the
wall.  Therefore, a soliton of finite dimensions requires a domain
wall with an inhomogeneity of some sort, e.g., a wiggle.  Here we will
assume that the wavefunction of a zero mode decays quickly enough in
all directions.

Zero fermion modes always come in doublets. This is essentially a
consequence of the charge conjugation symmetry (\ref{symm-sigma2}) at
$E=0$.  More formally, by starting with the symmetries of zero modes
(\ref{symm-sigma2}) and (\ref{CP}), we can construct a triplet of
SU(2) generators
\beq
S_2 = \frac{\sigma_2}{2},
\ 
S_3 = \frac{\sigma_3 e^{i\bf Q\cdot r}}{2},
\ 
S_1 = -i[S_2,S_3] = \frac{\sigma_1 e^{i\bf Q\cdot r}}{2}.
\label{modified-spin}
\eeq
Components of $\vec{S}$ are conserved quantities for a zero
mode.  By inspection, $\vec{S}\!\cdot\!\vec{S} = 3/4$, i.e., a zero
mode is a doublet ($S=1/2$).  Physically, $S_3$ is the component of
the total spin of the system parallel to staggered magnetization and
is therefore a conserved quantum number.  $S_1$ and $S_2$ are
components of the total {\em staggered} spin.  They generate {\em
staggered} rotations of spins and are conserved for zero mode fermions,
but not for bulk states.  

\subsection{Spinons and holons}

We consider a system with only one zero-mode doublet localized on a
topological defect at the origin, ${\bf r}=0$.  We assume that the
system is symmetric, i.e., $m(-{\bf r}) = -m({\bf r})$.

{\em Half-filled system.}  First let us show that a half-filled system
has a uniform density.  States with $E<0$ are filled, while those with
$E>0$ are empty.  One of the two zero modes is filled.  Because the
density contribution of the zero mode $\psi_0^\dagger({\bf
r})\psi_0({\bf r})$ is invariant under rotations generated by
$\vec{S}$, we can consider the situation when the mode with $S_3=1/2$
is occupied without loss of generality.  Then the operator $\sigma_2$
toggles between occupied and unoccupied states.  The expectation value
for the density is
\begin{eqnarray}
n({\bf r}) &=& \sum_\psi^{\rm occ} \psi^\dagger({\bf
r})\psi({\bf r}) = \sum_\psi^{\rm occ}
\psi^\dagger({\bf r})\sigma_2^2\psi({\bf r}) 
\nonumber\\ 
&=&
\sum_\psi^{\rm unocc} \psi^\dagger({\bf
r})\psi({\bf r}) 
= {1\over2}\sum_\psi^{\rm all}
\psi^\dagger({\bf r})\psi({\bf r}) = 1.
\nonumber
\end{eqnarray}
Thus a half-filled system has a uniform charge density.  It will be shown
below that it contains a charge-0, spin-$1/2$ soliton at its center.  

{\em Half-filled system $\pm1$ electron.}
When a single electron or hole is added, the density distribution
$n({\bf r})$ is determined by the profile of the zero mode
$\psi_0^\dagger({\bf r})\psi_0({\bf r})$.  Because the wavefunction is 
localized around a defect at the center, we find a soliton with 
charge $\pm 1$.  

At the same time, the soliton has zero net spin $S_3$.  Moreover, 
$S_3=0$ in {\em any} symmetric finite area around the
defect.  This happens because contributions to the total spin from 
${\bf r}$ and $-{\bf r}$ cancel each other:
\[
\langle s_3({-\bf r})\rangle = (-1)^{x+y}m({-\bf r})  \!
= -(-1)^{x+y}m({\bf r})\!
= -\langle s_3({\bf r})\rangle.
\]

Thus, at half-filling $\pm 1$ electron, the system has a soliton
with charge $Q=\mp 1$ and spin $S_3 = 0$.  Exactly at half-filling, 
the soliton has $Q=0$ and $S_3 = \pm1/2$.  These are, respectively, 
a holon and a spinon.  

\section{Straight domain wall: a lattice analysis}
\label{straight-wall-app}

Consider a straight domain wall along the $x$ axis.  At a given
lattice momentum $k_x$ along the wall, the mean-field Hamiltonian
(\ref{HF-3-symm}) reads \begin{eqnarray} -\sigma_1 t [\psi(y+1) + \psi(y-1) +
2\cos{k_x}\psi(y)] \nonumber\\ - \sigma_3 U m(y)\, \psi(y) = E\,
\psi(y).
\label{HF-3-kx}
\end{eqnarray}
For definiteness, the domain wall is located on the line $y=0$, so
that $m(-y) = -m(y)$.  The site indices are integer ($0,\pm1,\ldots$)
for a {\em site}-centered wall and half-integer
($\pm1/2,\pm3/2,\ldots$) for a {\em bond}-centered wall.

We will first show that the smaller gap (separating the midgap bands) is
absent in the case of a {\em site}-centered domain wall, for which
$m(-y) = -m(y)$.  It suffices to show that there exists a zero mode at
$k_x=\pm\pi/2$.

\subsection{Site-centered wall: zero modes}

Setting $E=0$ and $k_x=\pm\pi/2$ converts Eq.~(\ref{HF-3-symm}) into
\beq
\psi(y+1) + \psi(y-1) = i\sigma_2 \Delta(y)\, \psi(y),
\label{0-mode-straight}
\eeq
where $\Delta(y) = U m(y)/t$. In the bulk, 
\[
\Delta(+\infty) = -\Delta(-\infty) = \Delta_0 > 0.
\] 
We are looking for a finite-norm solution to Eq.~(\ref{0-mode-straight}).
Evidently, $\sigma_2$ can be immediately diagonalized: $\sigma_2=\pm1$. 
The substitution 
\[
\psi(y) = \phi(y) e^{-i(\pi/2)\sigma_2 y}\, \chi
\]
where $\chi$ is an arbitrary constant spinor, yields a difference
equation with real coefficients for a scalar $\phi(y)$:
\beq
\phi(y+1) - \phi(y-1) = -\Delta(y) \phi(y).
\label{real-eqn}
\eeq

Eq.~(\ref{real-eqn}) has two linearly independent solutions with the 
asymptotic behaviors 
\[
\phi_\pm(y) \sim \exp{(\pm \kappa y)}
\hskip 5mm 
\mbox{ as } y\to +\infty. 
\]
The spread of the wavefunction $\kappa^{-1}$ is given by the equation 
\[2\sinh\kappa = \Delta_0.\] 
Clearly only $\phi_-(y)$ can be normalizable --- provided that it is
also well behaved as $y\to -\infty$.  It will be seen shortly that
this is indeed the case.

Observe that one can write $\phi_-(y)$ as a superposition of another
pair of linearly independent solutions, the eigenstates of parity,
\[\phi_{i}(-y) = \eta_{i}\phi_i(y),\] 
where $i = 1,2$ and $\eta_i = \pm 1$.  Ordinarily,
$\eta_1 \neq \eta_2$, so that
\[\phi_-(y) = c_1 \phi_1(y) + c_2 \phi_2(y)\]
is not, in general, a parity eigenstate.  Yet it {\em must} be if it is
to have a finite norm.  Indeed, from different asymptotic forms
of the solutions $\phi_\pm(y)$ we infer that $\phi_-(-y) = \eta
\phi_-(y)$, or else it will diverge at $y=-\infty$.  Luckily, 
{\em all} solutions of Eq.~(\ref{real-eqn}) have the same (even) parity
for a {\em site}-centered domain wall: $\Delta(0) = 0$ and therefore
\[
\phi(1) - \phi(-1) = -\Delta(0)\phi(0) = 0.
\]
Since $\phi_-(y)$ vanishes exponentially at both $y=+\infty$ and
$-\infty$, it has a finite norm.  

We have thus proven that the HF Hamiltonian (\ref{HF-3-kx})
has 2 solutions of a finite norm with $E=0$ for $|k_x|=\pi/2$, one for
each eigenvalue of $\sigma_2$, if the wall is {\em site}-centered.  As
a rule, there are no zero modes if the domain wall is {\em
bond}-centered.

The asymptotic behavior of the zero modes as $y\to\infty$ is 
\beq \psi({\bf r}) \sim
\exp{\left(\pm i \frac{\pi}{2}x -i\frac{\pi}{2}\sigma_2 y - \kappa|y|\right)} 
\, \chi, 
\eeq 
where $2\sinh\kappa = \Delta_0$ and $\chi$ is an arbitrary constant
spinor.  We stress that, in the limit of weak coupling $\kappa \ll
1$, the Fourier components of $\psi({\bf r})$ come from the 4 ``Fermi
patches'' $|k_x| \approx |k_y| \approx \pi/2$.  This means {\em twice
as many} Fermi points as in an ordinary 1D electron gas!  It is for
this reason that holons (and spinons) on a domain wall come in two
flavors.

\subsection{Site-centered wall: 1D electron band} 

For $k_x\neq\pi/2$, the degeneracy of the two midgap states is lifted
by the additional term $-2t\sigma_1\cos{k_x}$ in the HF Hamiltonian
(\ref{HF-3-kx}).  Near $k_x=\pm\pi/2$, this term can be treated as a
perturbation acting in the Hilbert space of the two zero modes,
allowing us to determine their splitting to first order.

Parametrize the wavefunctions in the familiar way,
\[
\psi(x,y) = e^{i k_x x}
\phi(y) e^{-i(\pi/2)\sigma_2 y}\, \chi,
\]
where $\phi(y)$ is the solution of Eq.~(\ref{real-eqn}) with the norm
1.  The resulting $\psi(x,y)$ diagonalizes the mean-field Hamiltonian
(\ref{HF-3-kx}) with $\cos{k_x}=0$.  The two-fold degeneracy of the
zero mode is due to the freedom in choosing the spinor $\chi$.  In the
framework of the degenerate perturbation theory, we compute the matrix
element of the perturbation $H_1 = - 2t\sigma_1\cos{k_x}$ between the
states $|k_{x1},\chi_1\rangle$ and $|k_{x2},\chi_2\rangle$: \begin{eqnarray}
\langle k_{x2},\chi_2|H_1|k_{x1},\chi_1\rangle = -2t\cos{k_{x1}}
\nonumber\\ \times \ 2\pi\delta(k_{x1}-k_{x2}) \
(\chi_2^\dagger\sigma_1\chi_1) \ \sum_{y}(-1)^y \phi^2(y).  \end{eqnarray}

This result is quite tangible: as far as the midgap states are
concerned, ``integrating out'' the transversal degree of freedom $y$
yields electrons with nearest-neighbor hopping along
the domain wall and no staggered magnetization ---
cf. Eq.~(\ref{HF-3-symm}):
\beq
-\sigma_1 t_{\rm 1D}[\psi(x+1) + \psi(x-1)] = E\psi(x).
\eeq
The effective 1D hopping amplitude is 
\beq t_{\rm 1D} = t\, \langle(-1)^y\rangle 
\equiv t\, \sum_{y}(-1)^y \phi^2(y).
\label{t1D}
\eeq
Note that $t_{\rm 1D} \ll t$ in the limit of weak coupling.


\begin{references}
\bibitem{0-pi} Note that lattice momenta of the kinks cover only a
half of the Brillouin zone, e.g., $0<k_x<\pi$.  This happens because
a state with a kink on site $n+1$ cannot be obtained from a state with
a kink on site $n$ by a mere translation of the system.  The minimal 
distance for such a translation would be 2 lattice spacings.
\bibitem{endnote:trugman} Hole motion by means of self-retracing
Trugman loops or by combining hops with spin exchanges is strongly
suppressed in the $J_z\to\infty$ limit.
\bibitem{endnote:collinear} We have also simulated the untwisting of
partially doped stripes when all components of the spin magnetisation
are present.  The untwisting first develops in a portion of the domain
wall containing no holes.  As it proceeds, holes are pushed away from
this region, letting the remaining undoped portion of the domain wall
to untwist completely, while the holes form the usual fully packed
stripes.  As a result, the remaining portions of the stripe have
semi-vortices attached at the ends.
\bibitem{neglect-rho} We have kept the density term intact in
numerical HF calculations of Sec.~\ref{Hubbard-numerics}.  Dropping it
in the subsequent discussion of the results in
Sec.~\ref{Hubbard-analytics} is both practical (at weak coupling, the
effects are spin-driven) and aesthetically pleasing (a charge
conjugation symmery is restored); see Ref.~\CITE{Schulz}.  In any
event, the identification of holons and spinons in numerical solutions
of the HF equations with the density term is unambiguous.
\bibitem{Bloch} This actually holds to any order in $m_{10}$, which
  can be proven using the analogy with precession of a spin 1/2 in a
  time-dependent magnetic field $\vec{B}(t) = -\vec{B}(-t)$.  If the
  expression in brackets, Eq.~(\ref{eq_0_modes}), is an odd function
  of $y$, an eigenstate of $\tau^y_3$ at $y = -\infty$ evolves into
  the same eigenstate of $\tau^y_3$ as $y\to+\infty$.  Then a solution
  $\psi(y)$ with $\sigma_2\tau^y_3 = - {\rm sgn}\, m_{11}(y)$ at
  $y=+\infty$ has a finite norm.
\end{references}
\end{document}